\documentclass[journal=jpccck,manuscript=article]{achemso}

\usepackage{graphicx}
\usepackage{dcolumn}
\usepackage{bm}
\usepackage{textcomp}

\setkeys{acs}{usetitle = true}

\usepackage{amsmath} 
\usepackage{amssymb} 
 \usepackage{amsfonts}

\usepackage{array}

\usepackage{multirow}

\usepackage{natbib}
\usepackage{mciteplus}

\author{Karol Palczynski}
\affiliation{Soft Matter and Functional Materials, Helmholtz-Zentrum Berlin, Hahn-Meitner Platz 1, 14109 Berlin, Germany}
\alsoaffiliation{Institut f{\"u}r Physik, Humboldt-Universit{\"at} zu Berlin, Newtonstr. 15, 12489 Berlin, Germany}

\author{Joachim Dzubiella}
\affiliation{Soft Matter and Functional Materials, Helmholtz-Zentrum Berlin, Hahn-Meitner Platz 1, 14109 Berlin, Germany}
\alsoaffiliation{Institut f{\"u}r Physik, Humboldt-Universit{\"at} zu Berlin, Newtonstr. 15, 12489 Berlin, Germany}
\email{joachim.dzubiella@helmholtz-berlin.de}

\title{Anisotropic Electrostatic Friction of Organic Molecules on ZnO Surfaces}

\keywords{Theory, Surface Diffusion, Anisotropic Diffusion, Hybrid Systems, Molecular Stochastic Langevin Dynamics, Sexiphenyl, Hexaphenyl, ZnO}

\begin{document}

\begin{abstract}

We study the long-time self-diffusion of a single conjugated organic {\it para}-sexiphenyl ({\it p}-6P) molecule
physisorbed on the inorganic ZnO $\left(10\overline{1}0\right)$ surface by means of  all-atom molecular dynamics computer simulations. 
We find strongly anisotropic diffusion processes in which the diffusive motion along the polar [0001] direction of the surface can be  
many orders of magnitudes slower at relevant experimental temperatures than in the perpendicular direction.  
The observation can be rationalized by the underlying charge pattern of the electrostatically heterogeneous surface which 
imposes direction-dependent energy barriers to the motion of the molecule. Furthermore, the diffusive behavior is found
to be normal and Arrhenius-like, governed by thermally activated energy barrier crossings.  The detailed analysis of the 
underlying potential energy landscape shows, however, that in general the activation barriers cannot be estimated from
idealized zero-temperature trajectories but must include the conformational and positional excursion of the molecule along its pathway. 
Furthermore, the corresponding (Helmholtz) free energy barriers are significantly smaller than the pure energetic barriers
with implications on absolute rate prediction at experimentally relevant temperatures.   
Our findings suggest that  adequately engineered substrate charge patterns could be possibly harvested to select desired growth 
modes of  hybrid interfaces for optoelectronic device engineering.
  
\end{abstract}

\maketitle


\section{\label{sec:intro}Introduction}

A large number of recent experimental and theoretical works have demonstrated that the fusion of conjugated organic molecules (COMs) and 
inorganic semiconductors to form hybrid inorganic-organic systems (HIOS) has an enormous 
application potential for the design of optimized optoelectronic devices.~\cite{Blumstengel:NJP, Koch, Trevethan_b} 
The reason for the anticipated capabilities of HIOS is the newly emerging physical and electronic properties of the hybrid interface, combining the favorable features of 
the two individual materials in a single conjugate.  For a future rational design of those hybrids, 
it is now of fundamental interest to better understand and rationalize the 
nature of the hybrid interface molecular structure, which, in turn, defines the matrix 
embedding the electronic excitations and energy level alignments needed for applications. 

During interface formation in thin-film epitaxy of exemplary rod-like organic molecules,~\cite{Simbrunner}  the COMs typically heterogeneously physisorb and 
self-assemble into well ordered and oriented crystals on the substrate surface.  The molecular 
attachment to the surface, subsequent surface diffusion, nucleation and growth is understood to be mainly driven by physical 
substrate-molecule interaction processes whose mechanisms, however, are in general quite convoluted.
The complexity not only arises due to  the intrinsic atomistic roughness but the additionally occurring terraces, 
step edges, and impurities on the substrate surface.~\cite{Ehrlich_a,Zhang:Science, hlawacek:sience, teichert:jpcm, draxl:prb} 
In particular, in HIOS materials, where the inorganic substrate surfaces are typically strongly polar,  
novel challenges appear due to highly anisotropic electrostatic interactions.~\cite{DValle} 

A few important insights into electrostatically controlled assembly have been given for HIOS recently in systematic experimental studies of the deposition of the prototypical 
COM {\it para}-sexiphenyl ({\it p}-6P) on inorganic 
ZnO surfaces. Here, it was found, for instance, that {\it p}-6P favorably adsorbs in a flat-lying fashion on the 
ZnO $\left(10\overline{1}0\right)$ surface with the long molecular  axis (LMA) perpendicular to the 
polar [0001] direction of the surface.~\cite{blumstengel:pccp}  Based on a combination of 
first-principle and classical theoretical approaches, this result was rationalized 
by the fact  that the intrinsic electrostatic surface pattern of the strongly ionic ZnO crystal aligns 
the  attachment of the multipolar (essentially quadropolar) molecule.~\cite{henneberger:prl} 
It was additionally demonstrated that  a high energy barrier exists for the COM translation along the direction of  
the polar axis on the order of tenths of eV, while hardly such a barrier is present 
for the translation perpendicular to the polar [0001] direction, if the LMA is also oriented in the same direction.

The just described underlying electrostatic energy landscape defines an underlying template
for the molecules to attach in a pre-defined fashion and nucleate. Therefore, it enables a controlled engineering of 
both inorganic surfaces and organic molecules with adapted polarity towards property optimization. 
However, growth of thin films from COMs deposited from the gas phase is an intrinsically nonequilibrium phenomenon 
governed by a subtle competition between kinetics and thermodynamics.~\cite{Zhang:Science} 
Precise control of the nucleation and growth and thus of the properties of hybrid interfaces becomes possible 
only after an understanding of the first kinetic steps is achieved~\cite{Marcon}.  Given such a complex surface energy landscape,  
the natural question now arises, how does it influence the time scale of diffusion and with thus possibly 
the nucleation rates of the COMs? Can it impose anisotropic kinetic barriers which lead to anisotropic transport and nucleation 
features far from equilibrium? How is the diffusion, which precedes nucleation, exactly connected to the energy landscape? 
If these questions are understood, apart from {}``templating'' according to equilibrium design rules, possibly also  
kinetic principles could be introduced to obtain desired nucleation modes and anisotropic seeds 
from electrostatic charge patterning for a controlled HIOS interface growth. 

This contribution is routed to address some of the questions above by studying the anisotropic
diffusive behavior of a single COM on the ZnO $\left(10\overline{1}0\right)$ surface using atomistically
resolved force-field computer simulations. The virtue of these simulations is that they capture atomistic details
of the system, and the thermally governed dynamic processes in the latter can be integrated into the relevant diffusive timescales,  as demonstrated, e.g.,  for
alkanes on metal surfaces,~\cite{Fichthorn} single atoms on a model bcc crystal surface~\cite{Resende_a} or on MgO~\cite{DGao}, benzene on graphite \cite{Fouquet},  C$_{60}$ (fullerenes) 
and pentacene on pentacene crystal substrates,\cite{Cantrell} and organic molecules on an insulating ionic (KBr)~\cite{Such} or (TiO$_2$) surface.~\cite{Trevethan}  
Most of the studies focused on molecular hopping mechanisms and diffusion on terraces, along steps, and 
attachment to and detachment from terraces and islands, which are among the most fundamental 
atomistic processes in the early stages of thin-film growth.~\cite{Zhang:Science}
However, to the best of our  knowledge, electrostatically-governed diffusion processes were 
not studied in detail  by simulations before. 

In our MD simulations of HIOS, we indeed find strongly anisotropic diffusion constants with orders of
magnitude difference between the diffusion along and perpendicular to the  polar [0001] direction of the 
surface, respectively. The diffusion constants are quantified and we show that they are indeed governed by 
 electrostatic friction contributions. However, detailed analyses of the underlying energy landscape 
 demonstrate that conformational and small positional fluctuations of the COM significantly influence both the potential 
 energy landscape and the free energy landscape, with large implications for the prediction of absolute
 rate constants and their temperature behavior.  Once better understood, we believe that the observed 
anisotropic electrostatic friction could be harvested for the kinetic control of nucleation and growth of 
hybrid interfaces for optoelectronic device engineering. 

\section{\label{sec:methods}Methods}

\subsection{MD simulations}

Our atomistically-resolved systems are simulated using the Gromacs simulation
package~\cite{Hess2008} (version 4.5.5) in combination with the generalized 
Amber force field (GAFF) for organic molecules.~\cite{Wang2004} Besides the provided intramolecular interactions, 
the intermolecular potentials are modeled by Lennard-Jones (LJ) and Coulomb interactions to account for van der Waals and electrostatic interactions, respectively. The partial charges placed on the individual atoms 
of the {\it p}-6P molecule have been calculated using the Gaussian 09 software~\cite{g09_short} by employing the B3LYP functional with 
the cc-PVTZ basis set using the electrostatic potential fitting method (ESP). We recently showed that with this force-field (Hamiltonian) 
the molecules have single-molecule conformations consistent with quantum calculations and self-assemble into the correct experimental 
room-temperature bulk crystal structures of{ \it p}-6P.~\cite{CGD}

\begin{figure}
\includegraphics[width=9cm]{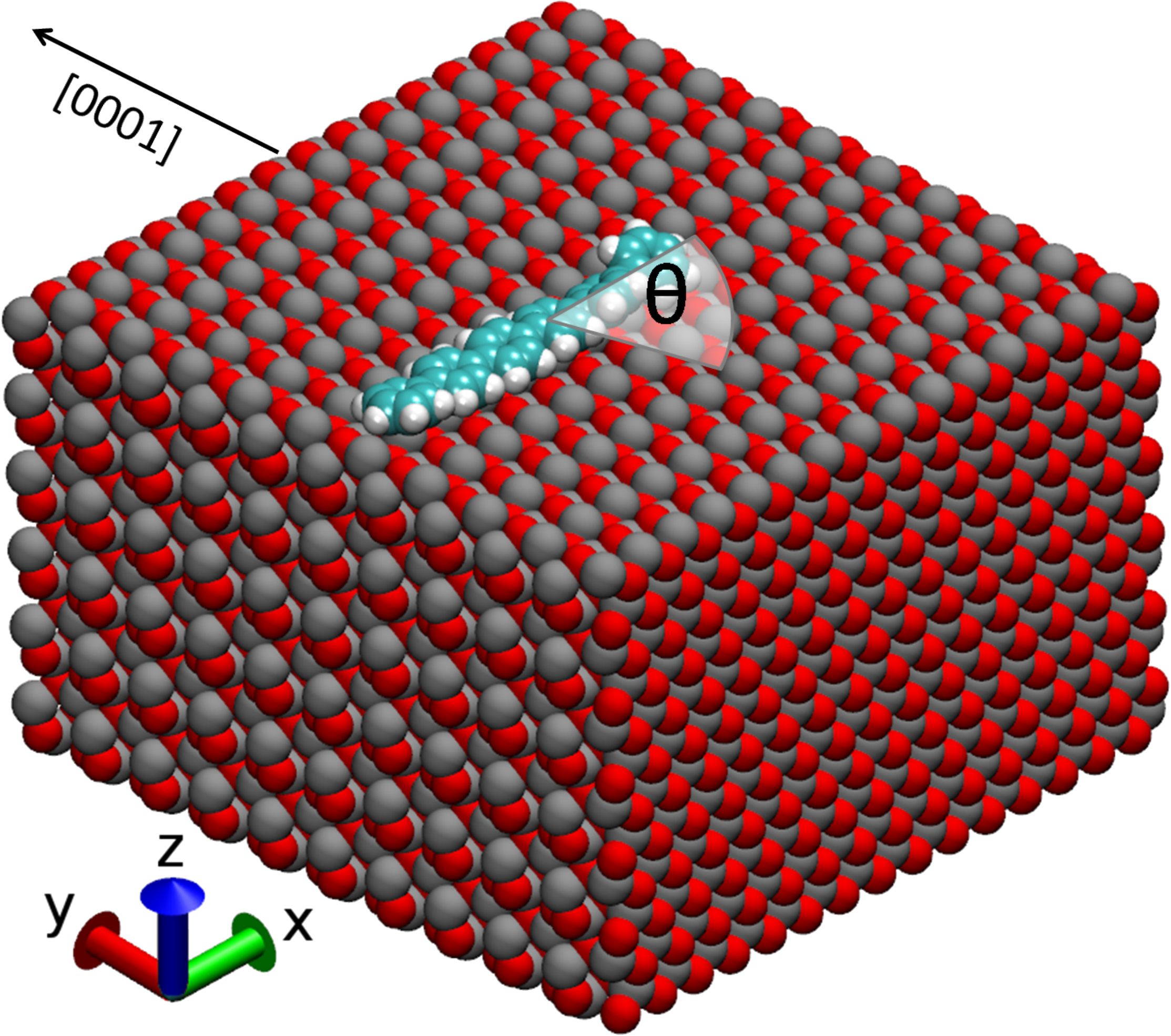}
\caption{Snapshot of the simulation system: it is comprised of a ZnO slab with the ZnO $\left(10\overline{1}0\right)$ terminated surface in $z$-direction and periodically repeated in $x$ and $y$ direction.  A single atomistically-resolved and fully flexible {\it p}-6P-molecule is adsorbed onto the surface and can freely diffuse, while the ZnO surface is frozen and acts as a static potential to the organic molecule. Due to the underlying charge pattern of the ZnO surface, the molecule is directed perpendicular to the (polar [0001]) $y$-direction with a high probability.~\cite{henneberger:prl} The variable $\theta$ denotes the angle between the LMA of the COM and the $x$-axis.}
\label{fig:setup}
\end{figure}

The ZnO surface is also atomistically resolved.  The atoms of the organic molecule interact with the ZnO surface via the LJ 
and Coulomb potentials. The LJ size and energy parameters of ZnO are taken from GAFF~\cite{Wang2004} and are $\sigma_{\rm Zn} = 1.96$~\AA, $\epsilon_{\rm Zn}=0.0523$~kJ/mol and $\sigma_{\rm O} = 2.96$~\AA, $\epsilon_{\rm O}=0.8764$~kJ/mol for the zinc and oxygen ion, respectively. The Lorentz-Berthelot mixing rules for the LJ cross interactions are applied.\cite{Hess2008}   The ZnO partial charges are taken from recent estimates~\cite{henneberger:prl} and  are $q_{\rm Zn}=0.95e$ and $q_{\rm O}=-0.95e$. Note that the latter stem from an empirical mapping of the solution of Poisson's equation to 
(also approximate) density-functional theory  results of the global electrostatic field and are therefore ill-defined. However, similar values as those employed 
($\pm 25\%$) are consistently found in literature,~\cite{zno1,zno2,zno3,zno4} and thus give the most reasonable classical 
representation of the charges.  
Explicit polarization effects of both the COM and the ZnO surface are neglected in our study. We thus focus
on the leading order static contribution of electrostatics to the diffusion process.  
We take the surface atoms frozen in time and space effectively resulting in a static surface potential for the COM. 
Our model system is comprised of a ZnO slab containing $N_{x}\times N_{y}\times N_{z}=15\times10\times6$ ZnO unit cells, periodically repeated in $x$ and $y$-directions with box lengths $L_x=4.935$~nm and $L_y=5.240$~nm, and a single {\it p}-6P-molecule  placed onto the ZnO $\left(10\overline{1}0\right)$ surface. 
A representative simulation snapshot depicting also the relevant directions is shown in Fig. \ref{fig:setup}. 
Note the highly structured surface pattern with the periodical repeats of the unit cell with 
wave lengths $l_x=0.329$~nm and $=l_y=0.524$~nm. The total number of atoms  simulated is $N=7262$. 

To avoid the energy-conservation and energy-partitioning problems typically faced in microcanonical 
($NVE$) simulations with constant energy $E$, we opt for stochastic MD simulations 
including auxiliary friction and noise terms. 
Such a Langevin approach warrants a quick dissipation and local equilibration and thus a  
smooth sampling of phase space for the calculation of the {\it long-time} dynamics. Since the 
imposed noise is uncorrelated to the real physical friction for long times, it can be properly subtracted afterwards 
(discussion follows). The underlying equation of motion for an atom $i$ 
at position $\vec r_i$ reads 
\begin{equation}
m_{i}\frac{\mathrm{d}^{2}\vec{r}_{i}}{\mathrm{d}t^{2}}=-m_{i}\xi_{i}\frac{\mathrm{d}\vec{r}_{i}}{\mathrm{d}t}+\vec F_{i}+{\vec{R}}_{i}\label{eq:SD}, 
\end{equation}
where $m_i$ is the atomic mass, $\xi_{i}$ is the friction constant, $\vec F_i$ the force acting on atom $i$, and ${\vec{R}}_{i}\left(t\right)$
is a Gaussian random force mimicking a white noise process obeying the fluctuation-dissipation theorem.~\cite{Hess2008}  The equations of motion are integrated using 
a leapfrog algorithm with a time-step of 2~fs.   All real space interactions including electrostatics are cut-off at at a radius of 1~nm. 
We find that this cut-off constitutes a good compromise between simulation speed and accuracy and is justified since there are
no long-ranged (monopole or dipolar) electrostatic interactions in the investigated system. 
The inverse friction constant $\tau_t$ is set to 2.0~ps, which ensures  both a constant average 
temperature and sufficiently large random micro-fluctuations. We have tested a range of reasonable 
friction constants between 0.02 and 20 ps and found, as expected, that 
the long-time (times $\gg \tau_t$) results for diffusion constant are independent 
of its particular choice. To study the dynamics of surface diffusion, the motion of the molecule is simulated in a set of 12 simulations each with a different temperature, ranging from 440~K to 820~K. Thanks to the small system size, the static surface, and the moderate cut-off distance, we were able to reproduce 1~ms of real-time dynamics.

\subsection{Derivation of energy landscapes}
 
In order to evaluate the energetic barriers in our simulations at zero temperature, we prepare an initial configuration with one molecule frozen in its
minimum configuration~\cite{CGD} on top of the ZnO surface with its LMA pointing in the $x$-direction, that is, 
the angle between the LMA of the COM and the $x$-direction is $\theta=0$ (cf. Fig. 1). 
The $z$-value of the center-of-mass coordinate of the COM (perpendicular
to the surface) is fixed to $z_{min}=0.3$~nm which constitutes the minimum potential 
state in $z$ as found in ab-initio calculations~\cite{henneberger:prl} and we do reproduce. 
We then scan the energy along the $x$ ($y$) directions for a fixed angle $\theta=0$, while the
 $y$ ($x$) position is chosen fixed in its minimum potential energy state, which we define to be $y=0$ ($x=0$). We also 
scan the energy resolved in the angle $\theta$ for fixed $x=0$ and $y=0$. This way of scanning the energy landscape is
the same as in the  {\it ab-initio} work~\cite{henneberger:prl} and constitutes the most reasonable one if the molecular
pathway is not known.  
For every coordinate $x$, $y$, and $\theta$, we produce a set of 500 configurations for one spatial period 
(that is, within the unit cell lengths $l_x$ or $l_y$ and 180$^\circ$, respectively). 
In each configuration we calculate the sum of LJ and Coulomb energies between the molecule and the surface. 

In order to calculate the 'real' energy landscape in directions $x$ and $y$ at elevated temperatures as well as estimate entropy contributions to the free
energy, we calculate the probability distribution $P(\alpha)$ of finding the COM center-of-mass at position 
$\alpha=x,y$ (modulus the wavelength of their period). The distribution in thus resolved in one direction, while the configurational excursions
in other directions are integrated out.  The free energy is 
then obtained from a standard Boltzmann inversion
\begin{eqnarray}
F(\alpha) = -k_BT \ln P(\alpha),  
\end{eqnarray}
where $\alpha=x,y$,  and the usual thermodynamic relation 
\begin{eqnarray}
F(\alpha) =  U(\alpha) - T S(\alpha)  
\end{eqnarray}
holds. Since we calculate the free energy at different temperatures $T$, the entropy can be obtained by the derivative
of $F(\alpha)$ with respect to $T$, $S = -\partial F/\partial T$ for the coordinate~$\alpha$. Numerically we calculate the derivative
by a simple finite-differences scheme $S(T) \simeq -[F(T+\Delta T) - F(T-\Delta T)] /2\Delta T$ with $\Delta T=25$~K. 
The energy follows directly as $E(\alpha)=F(\alpha)+TS(\alpha)$. 

\subsection{Long-time self diffusion constants}
The total one-dimensional long-time self diffusion coefficients $D_\alpha^{\rm tot}$ in $x$-direction ($y$-direction) are  formally 
obtained  from the $x$- ($y$-) component of the mean squared displacement (MSD) of the molecular center-of-mass, via 
\begin{eqnarray}
\left\langle \left(\alpha\left(t\right)-\alpha\left(t_0\right)\right)^{2}\right\rangle = \lim_{t\rightarrow\infty}2 D_\alpha^{\rm tot} t,  
\end{eqnarray}
with $\alpha =x,y$. 
In our simulations, there are two independent contributions to the final diffusion in the system.
We have a real, physical part contributing to the diffusion coming from the existence of static atoms which induce friction by the atomistic vdW and electrostatic interactions with the COM atoms. The frozen surface coordinates can be considered an average of the real-life atomic positions over long time-scales.
On the other hand we employ an auxiliary  random force in order to maintain a full energy dissipation and equipartition among the 
constituents. This contribution can be subtracted from the full friction: the total static friction constant $\xi^{\rm tot}$ is defined as the integral over the force-force
autocorrelation function 
\[\xi^{\rm tot}_\alpha=\lim_{\tau\rightarrow \infty}\frac{1}{3k_{B}T}\int_{0}^{\tau}\left\langle F_\alpha\left(0\right)F_\alpha\left(\tau'\right)\right\rangle {\rm d}\tau'\]
and the total force can be divided into a molecule-surface contribution,
$F^{ms}_\alpha$, and a molecule-bath part $F^{mb}_\alpha$. The latter auxiliary force is correlated to the molecular physics of the molecule-surface
system only on very small scale on the order of $\tau_t$. Consequently, the force cross-correlations $\left\langle F_\alpha^{mb}\left(0\right)F_\alpha^{ms}\left(\tau\right)\right\rangle +\left\langle F_\alpha^{mb}\left(\tau\right)F_\alpha^{ms}\left(0\right)\right\rangle $
vanish in the long-time limit, and the total friction is simply the sum 
 $\xi_\alpha^{\rm tot}=\xi_\alpha^{mb}+\xi_\alpha^{ms}$.  For our strongly interacting system we can safely 
 assume that we are in the high-friction regime~\cite{chen} and the usual Stokes-Einstein 
 (fluctuation-dissipation) relation, which reads 
 \begin{equation}
D_\alpha^{\rm tot}\left(T\right)=\frac{k_{B}T}{M\xi_\alpha^{\rm tot}}
\end{equation}
with the mass $M$ of the molecule and $k_BT$ the thermal energy. Hence,  the diffusion coefficient
can also be divided in two parts,
${1}/{D_\alpha^{\rm tot}\left(T\right)}={1}/{D_\alpha^{mb}\left(T\right)}+{1}/{D_\alpha^{ms}\left(T\right)}$,
and the desired molecule-bath diffusion in direction $\alpha$ calculated as 
\begin{eqnarray}
\frac{1}{D_\alpha \left(T\right)}:=\frac{1}{D_\alpha^{mb}\left(T\right)}=\frac{1}{D_\alpha^{\rm tot}\left(T\right)}-\frac{1}{D_\alpha^{ms}\left(T\right)}.
\end{eqnarray}

\section{\label{sec:results}Results and Discussion}

Our simulation results of the zero-temperature energetic 
potential of the center-of-mass coordinate of the COM in $x,y$-direction as well as  upon rotation $\theta$ versus direction $x$ 
are presented in Fig.~\ref{fig:Epot_y}. They qualitatively agree 
with the previous quantum DFT  calculations~\cite{henneberger:prl} but are quantitatively off by maximal 80\%.  Responsible for these deviations
are the approximations
in both methods, the quantum DFT as discussed in the previous work~\cite{henneberger:prl} as well as  
the MD simulations, for which the assignment of LJ parameters and partial charges to the ZnO surface is based on empirical mappings.   
However, all qualitative features rigorously agree between the various methods. In particular, the much stronger energetic corrugation in $y$-direction ($\Delta U_y \simeq 125$~kJ/mol)  than in $x$ ($\Delta U_x \simeq 1.3$~kJ/mol) 
suggests that at non-vanishing temperature, the molecule will diffuse significantly faster in $x$-direction with a weaker 
$T$-dependence. The angular corrugation suggests that it will do so in a highly directed fashion,  where the LMA 
favorably points into the $x$-direction. 

\begin{center}
\begin{figure}
\centering{}\includegraphics[width=9cm]{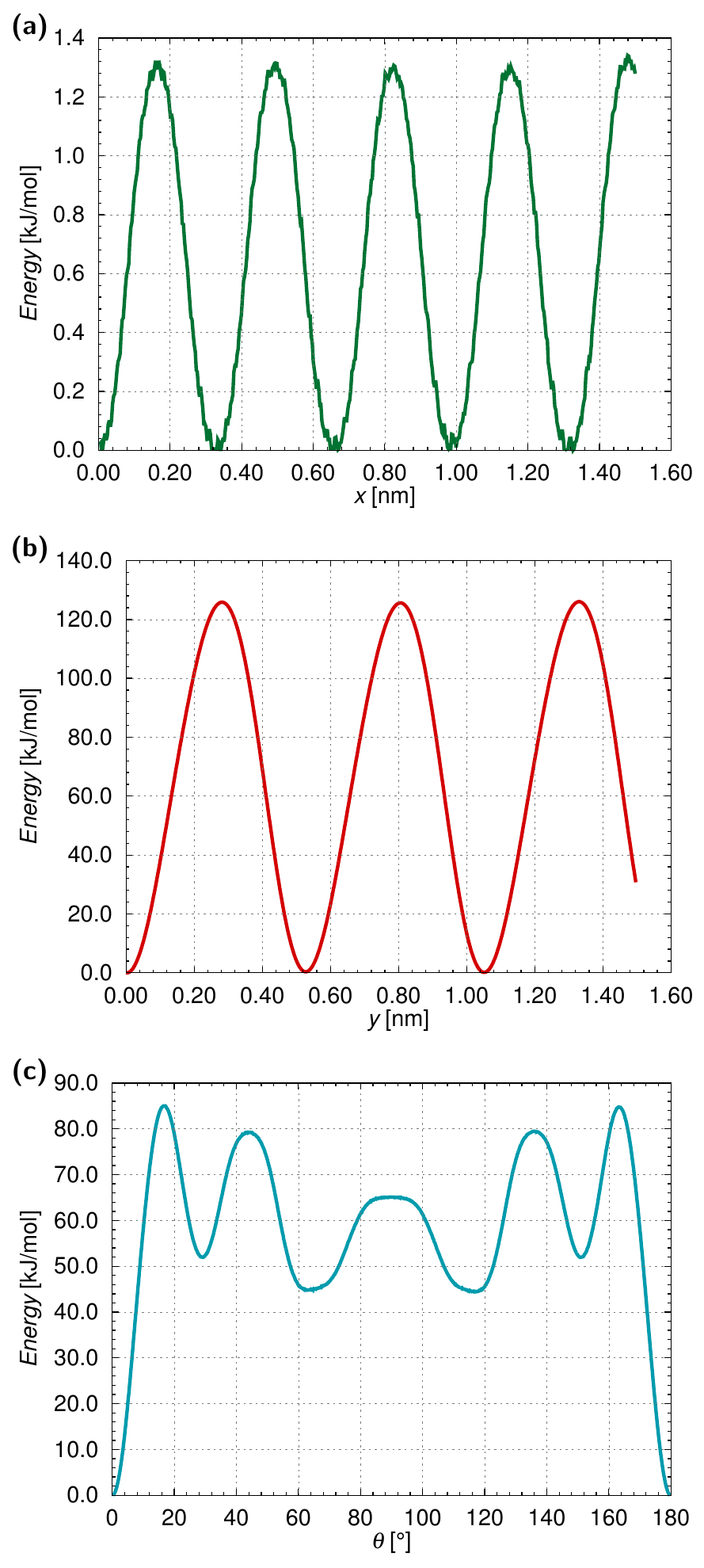}
\caption{Zero temperature energy landscape between the {\it p}-6P molecule and the ZnO surface in (a)  $x$-direction at $y=0$ and (b)  $y$-direction 
at $x=0$,  and (c) for the angle $\theta$ between the LMA of the COM and the $x$-direction at $x=0$ and $y=0$. }
\label{fig:Epot_y}
\end{figure}
\par\end{center}

Fig.~\ref{fig:pathways_charged} displays the real-space translational pathways the molecule takes on the surface over the course of simulations at temperatures
$T=440$~K, $T=670$~K, and $T=800$~K. It is indeed visible that at the lower investigated temperatures the motion in $y$-direction is significantly hampered in 
contrast to the motion in $x$-direction.  We find from the simulation trajectories that the {\it p}-6P molecule mostly slides along the rows of oxygen 
atoms, jumping, from time to time, across the potential energy  barriers in $y$-direction. At the highest  temperature (800 K), the jumps in $y$-direction appear much more often while the  preferred diffusion in $x$-direction is still clearly visible.  As already indicated in the snapshot in Fig.~\ref{fig:setup} and conjectured from  the energy surface, we indeed find that the organic molecule translates in $x$ in a directed fashion most of the 
time ($> 85\%$)  with its LMA pointing perpendicular to the (polar [0001]) $y$-direction within its variance. 
This is quantified in Fig.~(3), where  the average orientation distribution $P(\theta)$ strongly peaks at $\theta=0$ for all three temperatures.  
The  square root of the variance of the distribution is small and about  $\sqrt{\overline{\theta^{2}}}=2.8\pm2^\circ$.

\begin{center}
\begin{figure}
\centering{}\includegraphics[width=9cm]{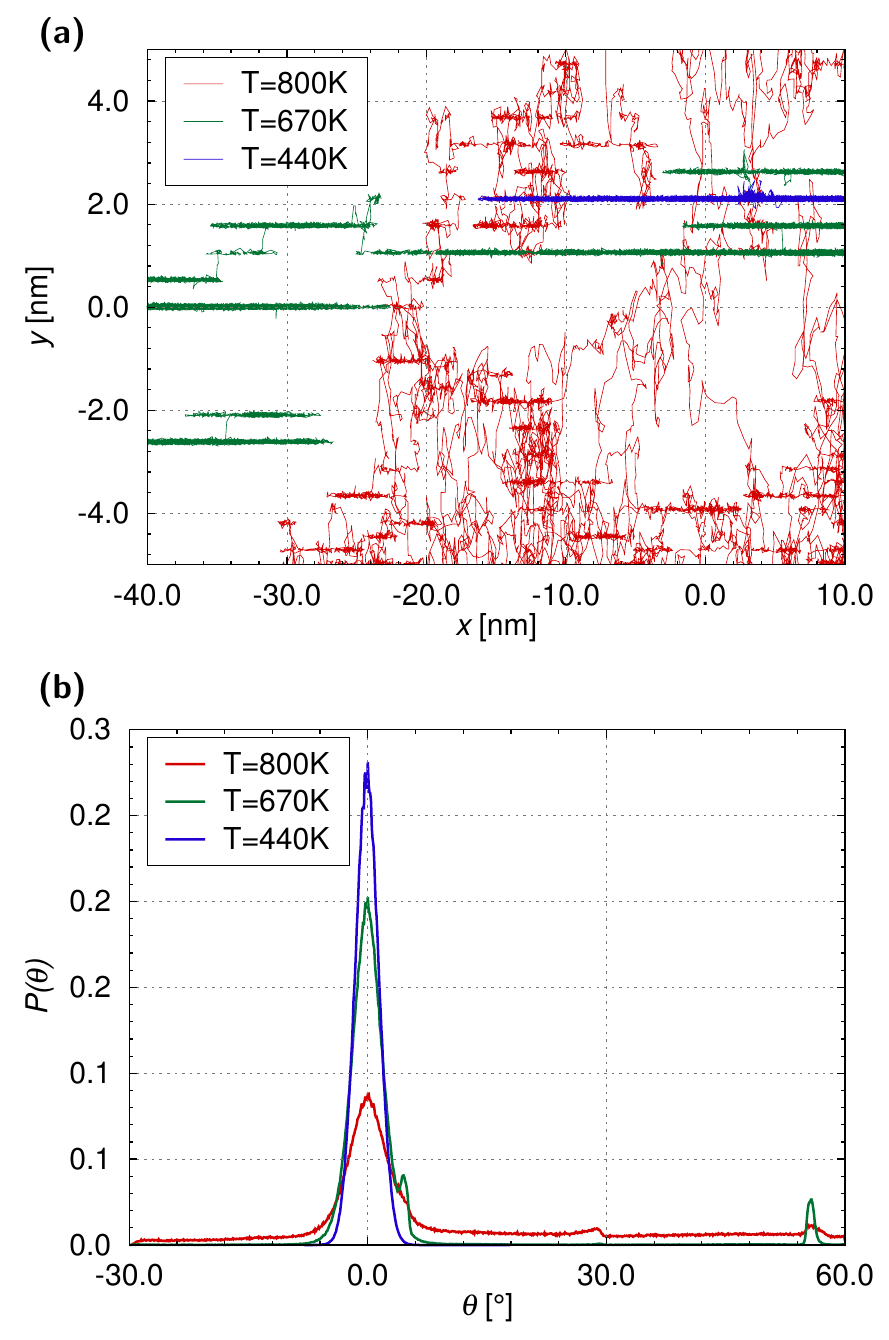}\caption{(a) Illustration of the real-space diffusion pathways of the 
{\it p-}6P molecule (center-of-mass motion) across the charged ZnO surface for three different temperatures as displayed in the legend. (b) Corresponding 
probability distribution  of the orientation $\theta$ of the LMA towards the $x$-direction.} \label{fig:pathways_charged}
\end{figure}
\par\end{center}

The calculated mean squared displacements (MSDs) are shown in Fig.~4 for both the $x$ and $y$-directions over more than two decades of time in the 
long-time limit ($t > 1$~ns).  In both cases the behavior is found to be mostly normally diffusive, that is, the MSD is proportional to $t^\beta$ with $\beta=1$.  In $y$-direction,  however, the slopes deteriorate for the lower temperatures ($T\lesssim 600$~K), indicating either sub-diffusive behavior or simply 
the lack of statistics because of the extremely slow dynamics (as discussed later).   From a linear fit of all normally behaving MSDs, we deduce the total  
long-time self-diffusion constants and calculate the wanted molecular-surface diffusion constants from eq.~(6). 
These $T$-dependent  molecular-surface diffusion constants are plotted in Fig.~(5) in an Arrhenius style, that is, the logarithm 
of $D_\alpha$ versus the inverse temperature  $1/T$.  As can be clearly seen, the diffusion coefficients display an 
extremely anisotropic dynamic behavior of the {\it p}-6P motion on the ZnO $\left(10\overline{1}0\right)$ surface.  
Only at the highest investigated temperatures ($>$ 800~K) , the magnitudes of the two diffusion constants are similar,  but 
already at roughly 600 K the diffusion in $y$-direction is about three orders of magnitude slower than in $x$!
As an example, in order to diffuse about  one nanometer in space at $T=590$~K, the COM needs  about 
a time of 0.1~ns in $x$, while it takes about 100 ns in $y$-direction. 

\begin{centering}
\begin{figure*}
\includegraphics[width=18cm]{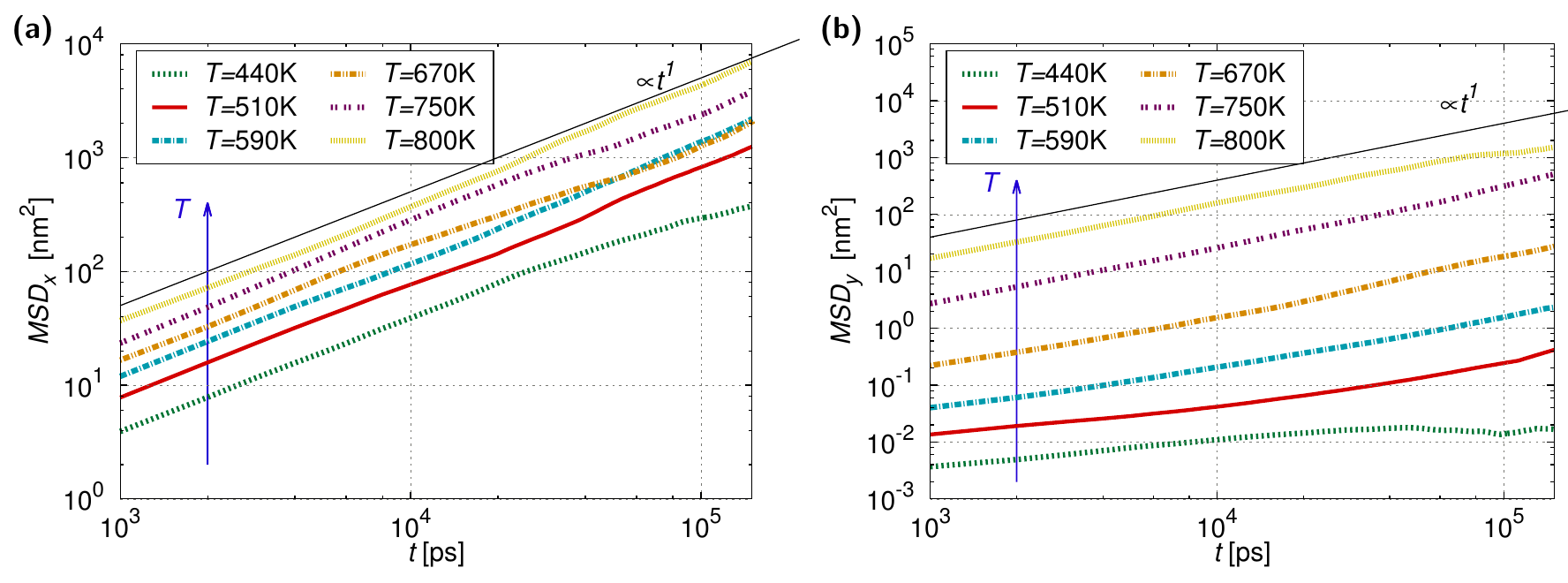}
\caption{The mean squared displacement (MSD) of the center-of-mass of the  {\it p}-6P molecule on the electrostatically charged ZnO $\left(10\overline{1}0\right)$ surface. 
(a) the MSD in $x$-direction; b) the MSD in $y$-direction, i.e., in the polar [0001] direction, cf. Fig. 1.}
\label{fig:msd}
\end{figure*}
\end{centering}

\begin{center}
\begin{figure}
\begin{centering}
\includegraphics[width=9cm]{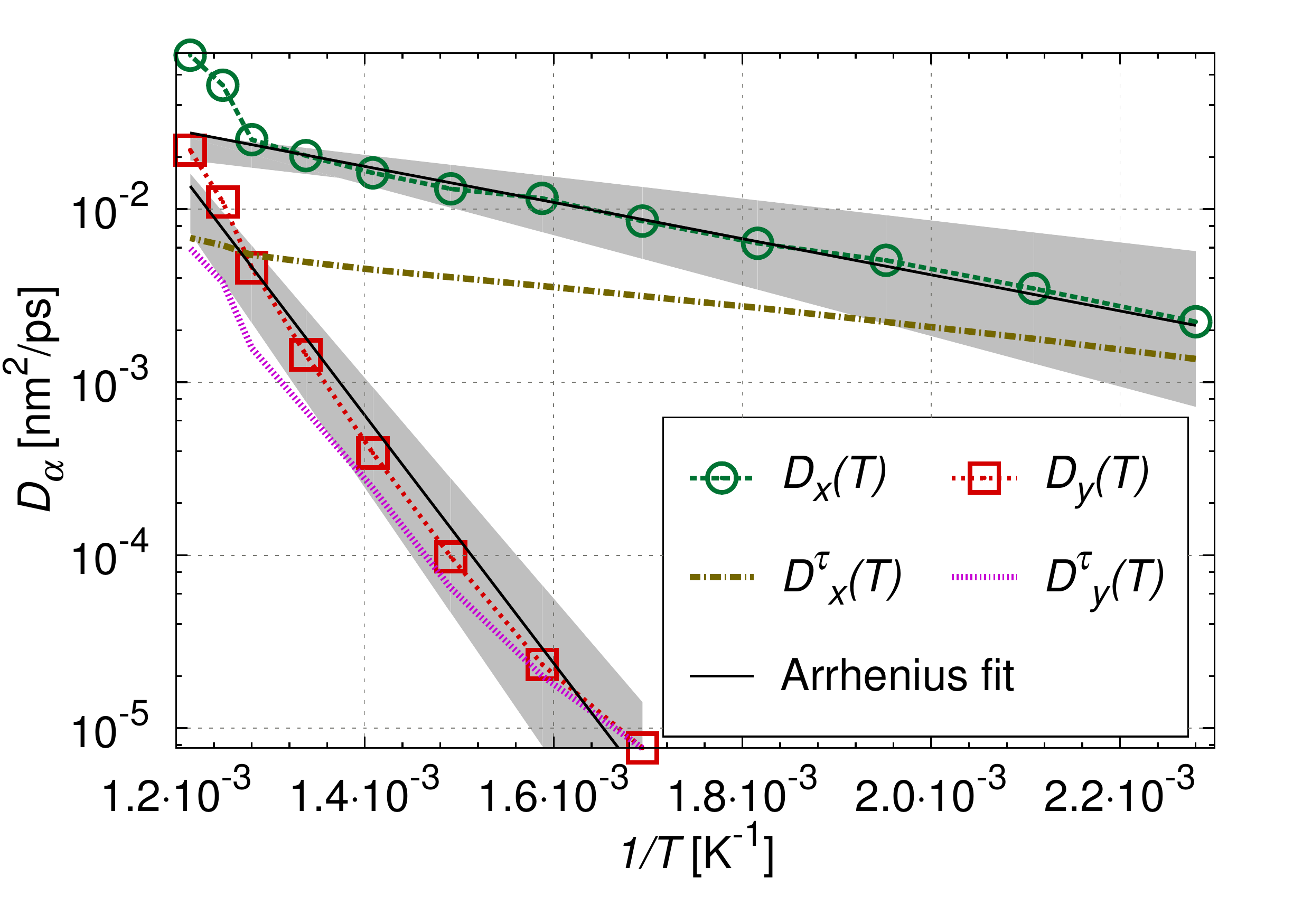}\caption{Simulation results (symbols) of the temperature dependent diffusion coefficients perpendicular to ($D_x$) and parallel  to ($D_y$) the polar $y$-direction. The curves can be nicely reproduced by a simple random-jump model (colored dashed lines and see text for description). From the linear fits (solid black lines with grey shaded error margins) in this Arrhenius-plot the effective energy barriers $\Delta U_\alpha$ can be deduced. \label{fig:D_T}}
\par\end{centering}
\end{figure}
\par\end{center}

Before interpreting the $T$-dependence of the data at hand of energy landscapes, we first show that the anisotropic diffusion is readily described by 
the mean waiting time in (or jumps between) the metastable states of the energetic potentials.  In this perspective,  the diffusion proceeds by 
uncorrelated jumps over the activation barriers at certain times between the adsorption potential wells in a well-defined periodic distance.
Hence,  the long-time over\-damped motion is characterized by the mean squared jump length $\left\langle l\right\rangle ^{2}$
and a mean waiting time $\tau$, i.e., the average time between two
consecutive jumps of length $l$. They can be related to the one-dimensional
$\left(\alpha=x,y\right)$ diffusion coefficient through~\cite{Ehrlich_a} 
\begin{equation}
D_{\alpha}^\tau=\frac{\left\langle l_{\alpha}\right\rangle ^{2}}{2\tau_{\alpha}}.
\label{eq:D_i}\end{equation}
The quantities $l_{\alpha}$ are in our case the surface lattice constants (wavelengths)  $l_{x}=0.329$~nm and $l_{y}=0.524$~nm.
The waiting times are calculated from the simulations by simply averaging the mean time the COM sits in a potential well before
a jump event. 

\begin{center}
\begin{figure}
\centering{}\includegraphics[width=9cm]{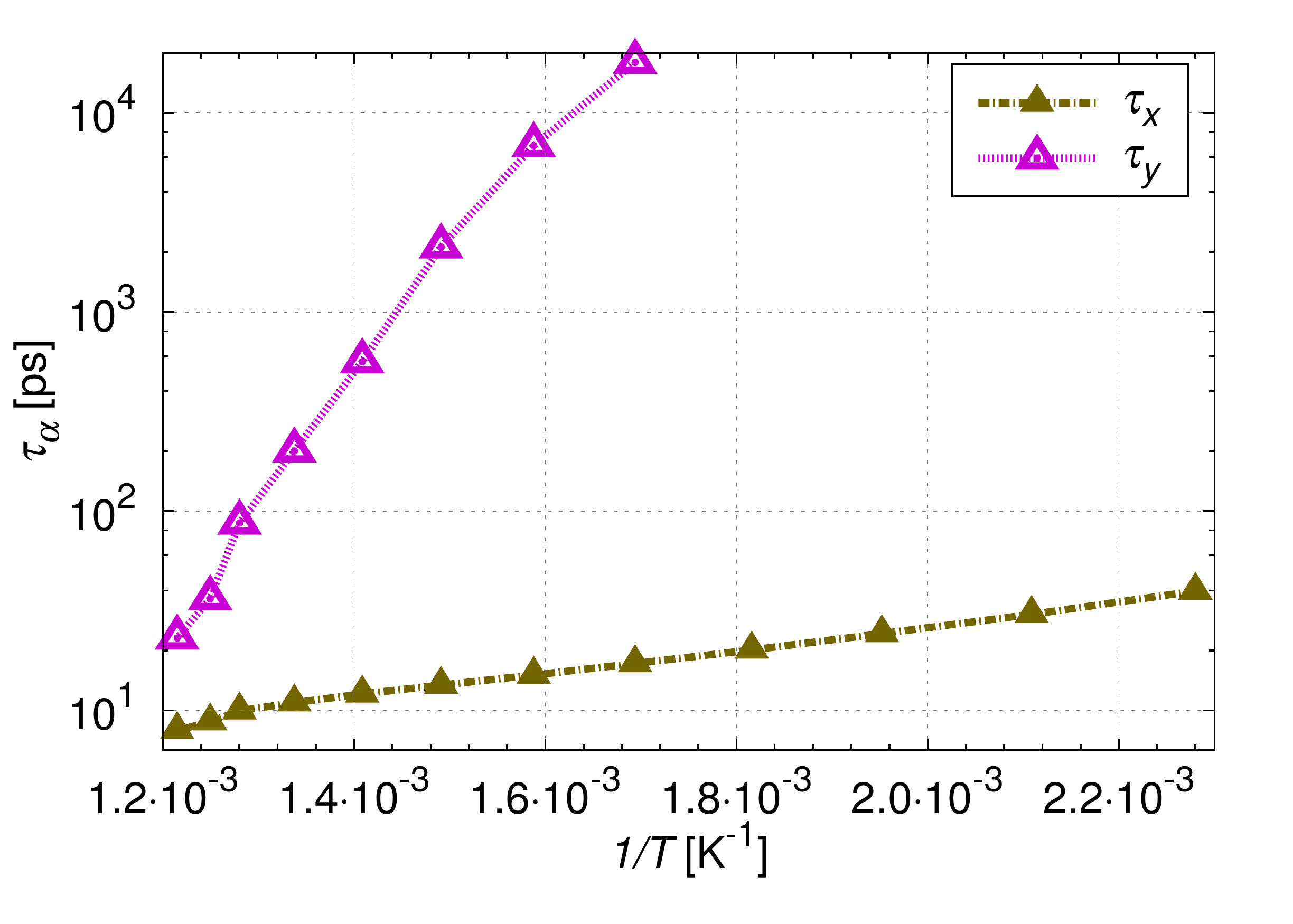}\caption{Mean waiting time for the jump from a potential well to a neighboring one 
as a function of temperature for directions perpendicular to ($\tau_x$) and parallel to ($\tau_y$) the polar $y$-direction. \label{fig:tau}}
\end{figure}
\par\end{center}

Fig. \ref{fig:tau} shows the mean waiting times in each of the one dimensions,
$\tau_{x}$ and $\tau_{y}$, in an Arrhenius-plot in the
range from $T=440$~K to $T=820$~K. For both  $\tau_{x}$ and $\tau_{y}$, 
we observe an almost linear growth with temperature and values 
under 40~ps in $x$-direction, while  $\tau_{y}$ on the other hand becomes extremely
high with peak values of 50~ns at $T=510$~K. At even lower temperatures,
jumps over the high potential barrier in $y$-direction occur only
once or twice during the entire 1~ms simulation, which causes a high
statistical error with inconclusive values (not shown).
The calculated values for $D_{x}^\tau$ and $D_{y}^\tau$ using eq.~\ref{eq:D_i}
are plotted in Fig.~\ref{fig:D_T} together with the diffusion coefficient derived from the MSD methods 
and show overall good agreement.  We can deduce from these fits that the long-time self diffusion of  a {\it p}-6P molecule 
is strictly governed by uncorrelated random jumps between the potential wells forming
lanes imposed by the atomic surface interaction pattern.

Consequently, the diffusion process can be treated as a thermally activated transport process~\cite{Ehrlich_a} and $D_\alpha$ 
takes the Arrhenius form
\begin{equation}
D_\alpha\left(T\right)\propto{\rm e}^{{-\Delta U_\alpha}/{k_{B}T}}, 
\end{equation}
with $\alpha=x,y$ and $\Delta U_\alpha$ denoting the respective activation energy. The latter is directly given by the slope in Fig.~\ref{fig:D_T} and amounts to a large $\Delta U_y=137\pm15$~kJ/mol in $y$-direction and $\Delta U_x=20\pm7.5$~kJ/mol in $x$-direction. Let us now compare these values to 
the zero temperature and 'real' energy landscapes in the system. 

Looking back at the '$T=0$' energy landscape in Fig. 2, we find that the energy barrier in $y$-direction is very close to the behavior
found from the Arrhenius fitting. In both cases the barrier is large and the values are comparable, $\Delta U_y=137$~kJ/mol for the investigated $T$ 
versus  $\Delta U_y=125$~kJ/mol for $T=0$.  In absolute terms, in $x$-direction the $T=0$  values are similarly different, where 
$\Delta U_y=20$~kJ/mol for the investigated $T$ versus  a small $\Delta U_y=1.3$~kJ/mol for $T=0$ ; they differ
by about 18.7 kJ/mol, which, however, in relative terms is substantial. Since the barrier magnitude is situated in the exponent
of the Arrhenius equation, even small changes on the order of a few $k_BT$ have substantial impact on the $T$-dependence of
the diffusion constant. Thus, the agreement in $x$-direction is not quantitative, while satisfactory in relative terms ($<10\%$) in $y$. 

However, the differences can be reconciled by looking at the really sampled  energy landscape in Fig.~7, which we plot together with the full (Helmholtz) free energy and the entropy. Evidently, the energy barriers are now very consistent with the ones  estimated from  the Arrhenius slopes: we find  $\Delta U_y=130\pm5$~kJ/mol  and $\Delta U_x=19.3\pm1$~kJ/mol.  The reason why the idealised $T=0$ energy landscape fails to describe the Arrhenius behavior quantitatively must be thus attributed to the 
idealized pathways of the COM in these calculations. In reality, under the influence of temperature, the COM motion is governed
by conformational and positional fluctuations which change the average interaction energies. Such a behavior was
observed before for functionalized organic truxenes on insulating KBr surfaces~\cite{Such} and large organic molecules with polar binding groups on the perfect TiO2 (110) surface.~\cite{Trevethan} In both studies detailed investigations by molecular simulations demonstrated that 
the diffusional pathway sensitively depends on the details of the molecular structure, such as flexibility and cooperative motions
of intramolecular groups. Interestingly, we find in our study that these 
excursions from the idealized pathways are small: during its motion along the surface in $x$, for instance,
 the standard deviation of the center-of-mass position in the
$y$-direction is less than 0.05 nm and in $\theta$ only less than 2.8\textdegree{}. We see in Fig.~7 that in both directions 
the fluctuations increase the height of the energy barriers which has in particular large consequences on the absolute barrier height
in $x$-direction. Another surprising issue is that the influence of these fluctuation on the average energy has only  a weak 
$T$-dependence, at least in the investigated $T$ range.  Otherwise, we would observe clear deviations from linearity 
(the Arrhenius behavior) in Fig. 5. 

Glancing back to the free energy profiles in Fig. 7, it is interesting to see that the entropy contribution to the free energy 
is substantial and almost cancels out the internal energy contributions. Overall the free energy barriers in both directions are much smaller
than the energy barriers. Since for quantitative (jump) rate predictions~\cite{kramer} for a fixed temperature the free energy
barrier is the decisive one, this finding has large implications for the interpretation of data and further theoretical 
modeling studies. For example, in $y$-direction the free energy barrier is about four times smaller than the pure energy barrier, 
so that in the former case a jump probability is about $e^4\approx55$ higher than in the latter case.  Hence, to properly interpret and 
describe transport processes of COMs on inorganic surfaces thus the full knowledge of the free energy landscape, 
including fluctuations, has to be available. 

\begin{figure}
\includegraphics[width=9cm]{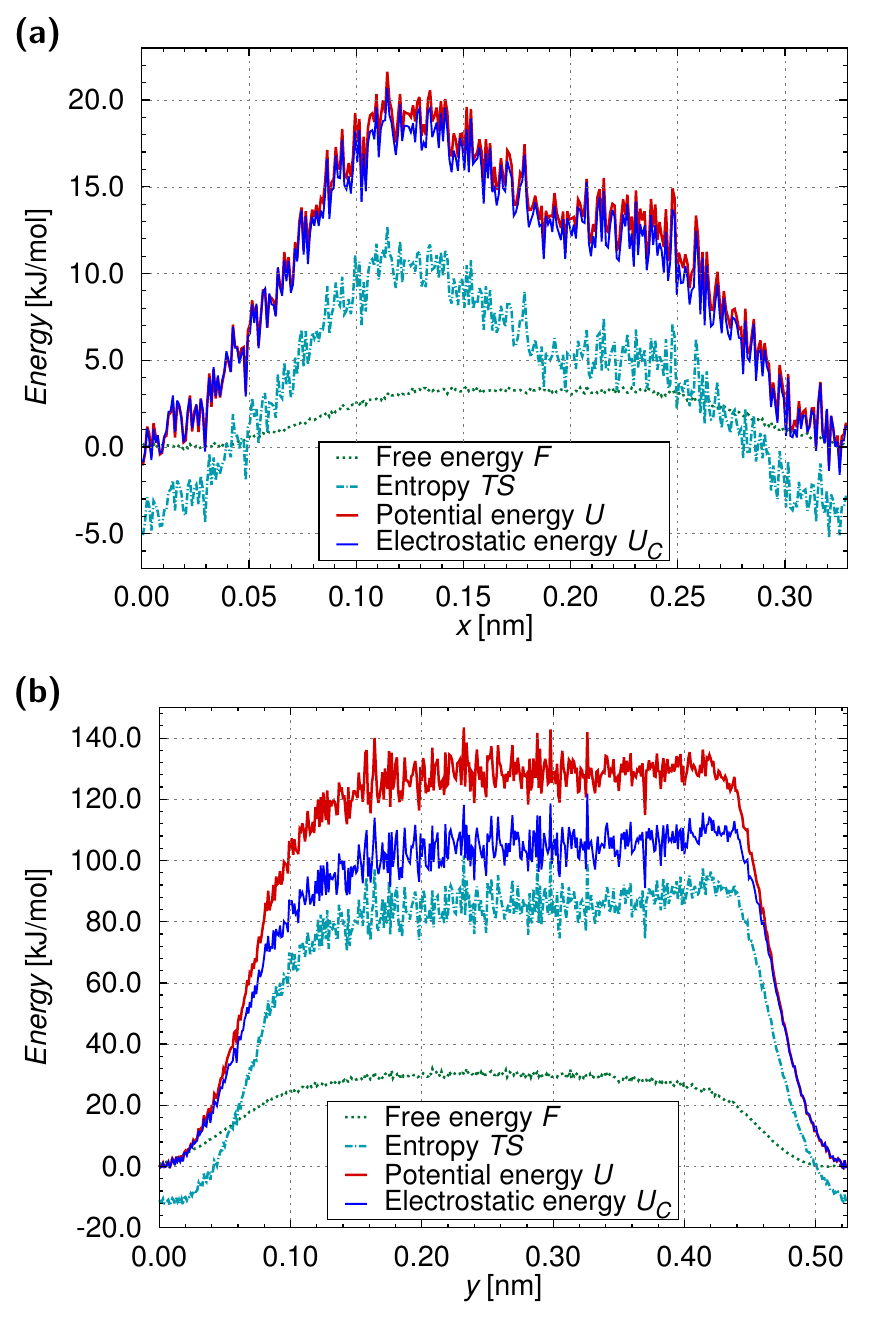}
\caption{Free energy $F(\alpha)=U(\alpha)-TS(\alpha)$, energy $U(\alpha)$, and the entropic contribution $TS(\alpha)$ resolved in direction $\alpha = x$ (a) and $\alpha = y$ (b) for  a temperature $T=723$~K. Also shown is the electrostatic surface-COM interaction part of the energy $U_C$. }
\label{fig:FU}
\end{figure}

\begin{figure}
\includegraphics[width=9cm]{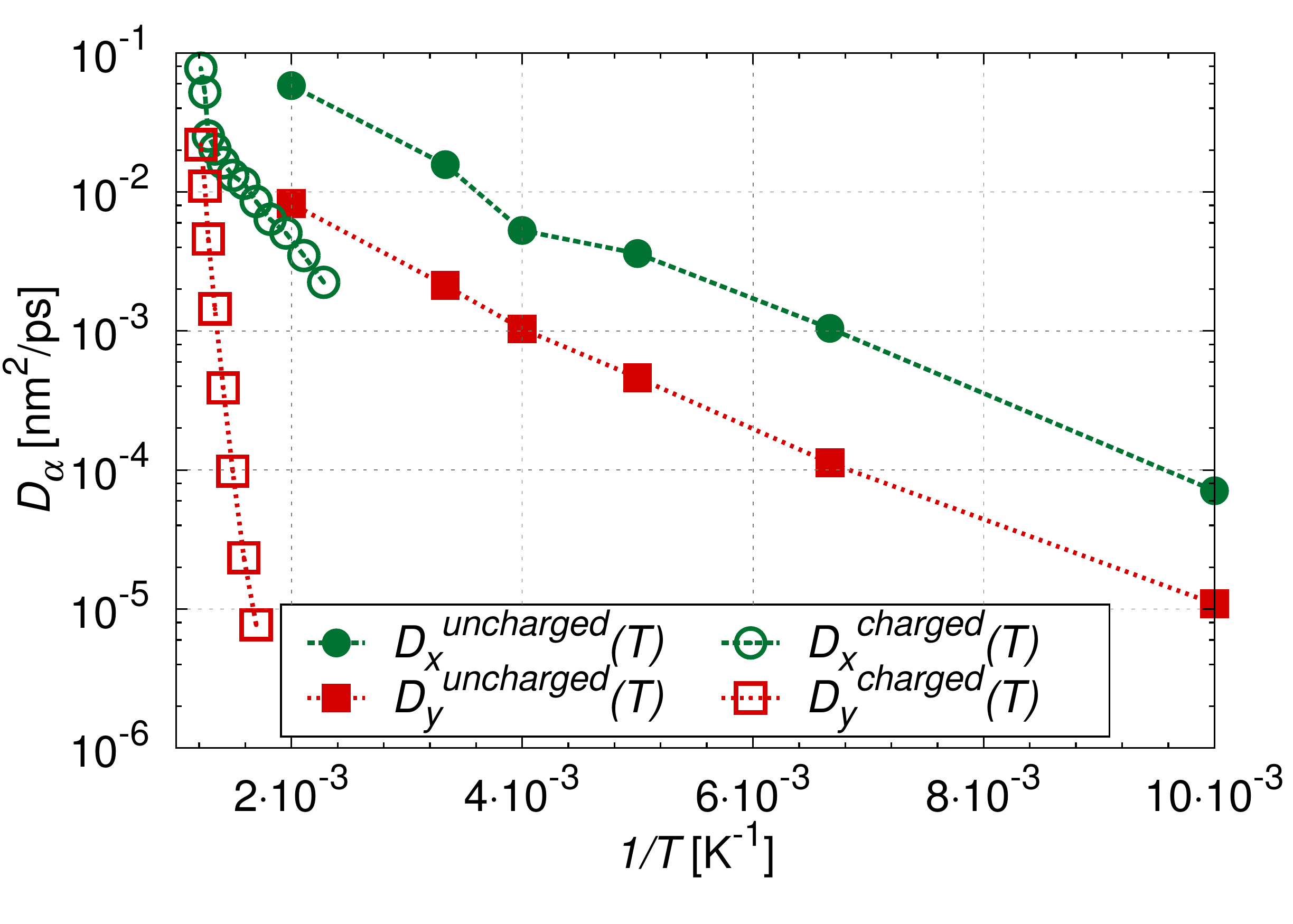}
\caption{Temperature dependent diffusion coefficients $D_\alpha(T)$ as a function of (inverse) temperature $1/T$ from simulations with all partial charges of the ZnO 
surface (not of the COM) switched off (full symbols) compared to the fully electrostatically coupled system (open symbols). }
\label{fig:D_q0}
\end{figure}

Finally, we also display the electrostatic energy part in Fig.~7. Clearly, and as anticipated, it contributes  at least 80\% to the total internal energy. 
The anisotropic long-time friction behavior is therefore dominated by electrostatics.  As a further consistency check, we have also simulated 
our {\it p}-6P/ZnO system with all the partial-charges in the ZnO atoms set to zero. Due to the weaker molecule-substrate interactions we have to shift the
 temperature-range down to 100 to 500~K to sample over adsorbed states. The results are summarized in Fig.~\ref{fig:D_q0}, compared to 
 the data of the full 
 electrostatically charged system in Fig.~5. Clearly, the $T$-dependence, and thus the activation barriers as well, are much smaller than for the fully coupled system. 
 The activation energies from fitting to the Arrhenius law for the neutral ZnO system are only about 6 kJ/mol equally for both directions. 
 Thus, the existence of partial charges on the surface imposes a strong inhomogeneity in the surface diffusivity and dominates the 
 long-time diffusion process.

\section{Summary and Concluding Remarks}

In summary, we have shown that the heterogeneous electrostatic surface pattern appearing on inorganic crystal
semiconductor surfaces leads to a strongly anisotropic surface long-time self-diffusion of organic molecules. The diffusive behavior is found to 
be normal -- within the investigated $T$-range where reasonable statistics could be gathered -- and is determined by thermally activated hopping between energy barriers. In our case of the {\it p}-6P COM diffusion on the ZnO $\left(10\overline{1}0\right)$ surface, this anisotropic electrostatic friction 
leads to a three orders of magnitude slower diffusion (for temperatures below 600~K) 
in one surface direction than in the perpendicular one. The found Arrhenius-like temperature behavior
 suggests an even more drastic difference for room temperature diffusion. 

The detailed analyses of the underlying potential energy landscape 
 demonstrate, however,  that thermal conformational and positional fluctuations of the COM significantly 
 influence the diffusion process as observed in related computational studies before.~\cite{Such, Trevethan} 
 In particular, we find that the potential energy barriers significantly deviate from those derived by zero K
 calculations of idealized pathways.  Only the 'real' energy landscape at the relevant temperature  
 for the fluctuating system can  quantitatively describe the $T$-dependence of the diffusion constants.
 Additionally, the free energy barriers at a fixed temperature deviates substantially from the 
 magnitude of the internal energy barriers.  This finding has large implications for the prediction of absolute
 rate constants~\cite{kramer} and their temperature behavior.  We believe that the observed anisotropic 
 electrostatic friction could be harvested for the kinetic control of nucleation and growth of 
hybrid interfaces for optoelectronic device engineering.  

\section{Acknowledgements}
This project was funded by the Deutsche Forschungsgmeinschaft (DFG) 
within the Collaborative Research Center 951 (SFB 951, project A1). 
The authors wish to thank Cemil Yigit for inspiring discussions. 

\bibliography{main}
\clearpage
\newpage

\begin{figure}
\includegraphics[width=18cm]{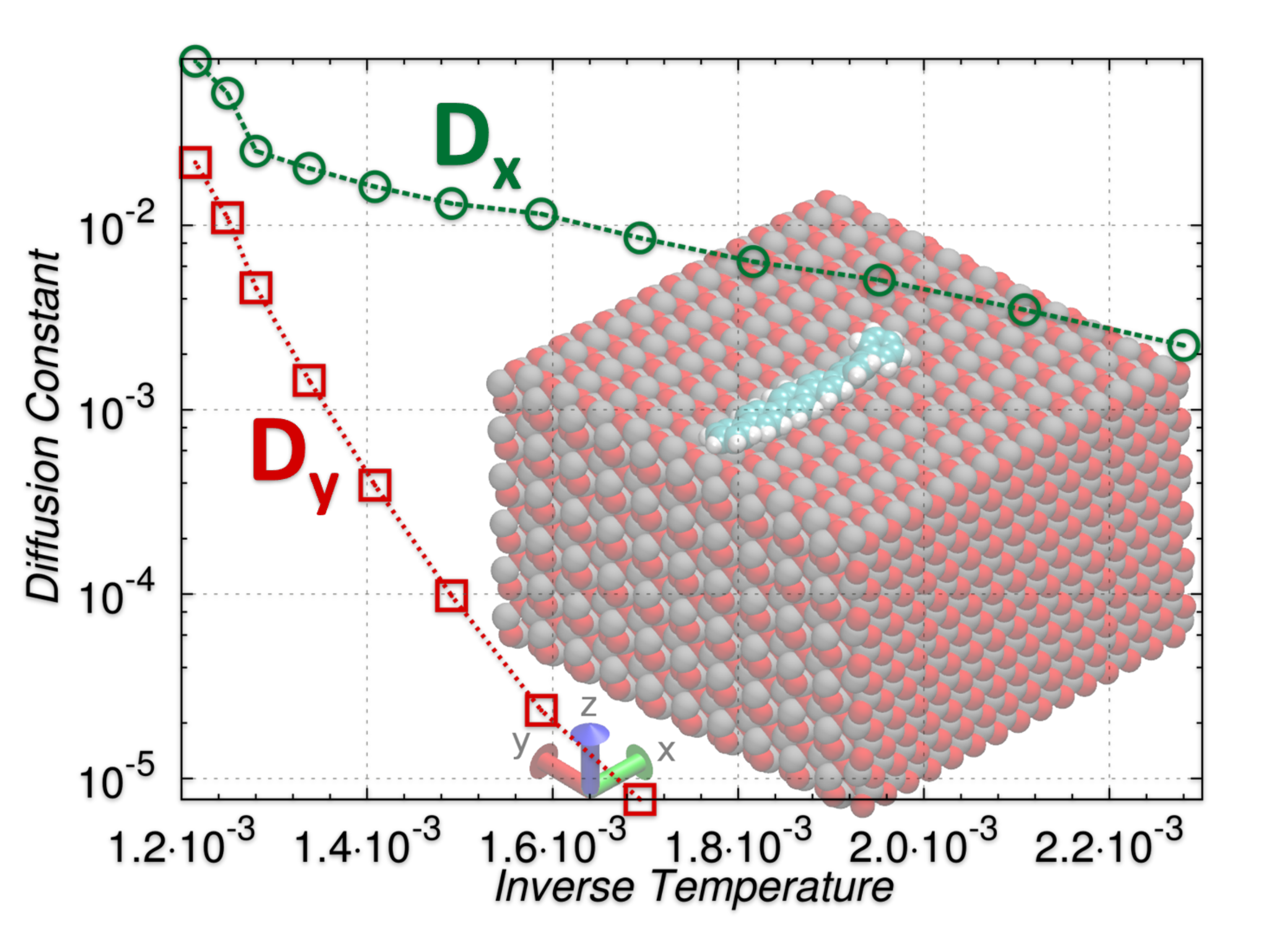}
\caption{Table of Contents Figure.}
\end{figure}

\end{document}